\documentclass[11pt,aps,superscriptaddress,nofootinbib]{revtex4}
\usepackage{graphics}
\usepackage{graphicx}
\usepackage{epsfig}

\newcommand{\bea}{\begin{eqnarray}}
\newcommand{\be}{\begin{equation}}
\newcommand{\eea}{\end{eqnarray}}
\newcommand{\ee}{\end{equation}}
\providecommand*{\ler}{\stackrel{\scriptstyle <}{\scriptstyle \sim}}

\begin{document}

\begin{flushright}
\preprint~CPHT-RR-024.0405 \\ ~LPT-ORSAY-05.27 \\~hep-th/0506172
\end{flushright}

\title{Large D-terms, hierarchical soft spectra and moduli stabilisation }

\author{Emilian Dudas}
\affiliation{Centre de Physique Theorique
\footnote{Unit{\'e} mixte du CNRS et de l'EP, UMR 7644.},
Ecole Polytechnique-CPHT, 91128 Palaiseau Cedex, France}
\affiliation{ LPT\footnote{Unit{\'e} mixte du CNRS, UMR 8627.},
B{\^a}t. 210, Univ. de Paris-Sud, F-91405 Orsay, France}

\author{Sudhir K Vempati}
\affiliation{Centre de Physique Theorique
\footnote{Unit{\'e} mixte du CNRS et de l'EP, UMR 7644.},
Ecole Polytechnique-CPHT, 91128 Palaiseau Cedex, France}

\begin{abstract}
We derive general expressions for soft terms in supergravity where
D-terms contribute significantly to the supersymmetry breaking in
addition to the standard F-type breaking terms. Such D-terms can 
strongly influence the scalar mass squared terms, while having
limited impact on gaugino masses and the B-terms. 
We present parameterisations for the soft terms when D-terms 
dominate over F-terms or become comparable with them. Novel
patterns emerge which can be tested phenomenologically. 
In a mixed anomaly-D mediated scenario, the scalars have masses 
from D-mediation, whereas gaugino masses are generated by anomaly 
mediation. 
As an application of this analysis, we show that while the 
"split supersymmetry" like mass spectrum with one fine tuned Higgs 
is not an automatic outcome of these scenarios, explicit models can
be constructed where it can be realised. 
 Finally, we show that large D-mediated
supersymmetry breaking can be realised in string  models
based on intersecting D-branes. Examples are presented where the moduli
are stabilised in the presence of large D-terms using non-perturbative
gaugino condensation like effects.
\end{abstract}

\maketitle

\section{Introduction}
In most models of supersymmetry (SUSY) breaking, supersymmetry is 
broken spontaneously in a hidden sector and is then transmitted 
to the visible sector through some interactions, mostly
gravitational. In supergravity, the hidden sector typically contains
a set of chiral fields whose auxiliary components attain a \textit{vev}
at the minimum breaking supersymmetry spontaneously. It is generally
preferred to have a dynamical explanation to this phenomenon. 
This breaking is communicated to the visible sector through tree
level (and higher order) gravitational interactions. After integrating out 
the heavy fields, including the hidden sector 
fields, the resulting effective lagrangian contains renormalisable supersymmetry 
breaking soft terms \cite{nilles,carlos}. At the full supergravity (SUGRA)
level, the soft terms are typically given in terms of the gauge kinetic 
function $f$, the K{\"a}hler potential $K$ and the superpotential $W$. 
Thus in the global limit, the structure of the soft terms crucially 
depends on the forms these functions take in SUGRA. 
For example, if the K{\"a}hler and the gauge kinetic functions are
canonical, this will lead to a universal soft spectrum with mSUGRA
boundary conditions. 

While analysis of the above type are suitable for simplest classes
of supersymmetry breaking models, for more complex situations it is useful
to have general expressions for soft terms \cite{weldonsoni,kaplunovskylouis}. 
Such situations can typically
arise when supersymmetry breaking has its origins in string theory. Given that
we do not yet have a concrete model of supersymmetry breaking in string theory,
it is much more advantageous to parameterise this breaking in terms of a few
parameters. In terms of effective supergravity lagrangians derived from string theory, 
the breaking can be parameterised as the \textit{vev}s of the auxiliary fields of the 
chiral superfields associated with the higher dimensional
gravitational multiplet, namely the dilaton field $S$ and the
moduli fields $T_i$, which effectively act as hidden sector fields. 
The main advantage of such parameterisations is that they could
capture the generic features of soft spectrum emanating from a class of
models without completely resorting to explicit model building. These features
could then be contrasted with the phenomenological requirements.  
Detailed analysis parametrising the resultant soft terms for the
heterotic case have been presented in \cite{ibanez}. Recently, they have
been further extended to the case of Type-I strings \cite{ibaneztypeI}.

The above analysis which has been very useful can however, be
considered as incomplete. This is because, they have
implicitly assumed only $F$-type breaking of supersymmetry (only
auxiliary fields of the chiral multiplets get a \textit{vev}). 
In a more generic scenario, it is well known that there could be D-type 
susy breaking contributions too \cite{dps}. These can arise for example in 
models based on anomalous $U(1)$ symmetries \cite{bd}. Furthermore, 
in effective lagrangians from the Type II orientifolds
with intersecting D-branes, one can expect such D-term contributions
to be naturally present. Given these motivations, it is natural to
extend the previous analysis by considering D-type SUSY breaking sources. 
In section two, we present these general expressions of 
soft terms, initially for generic fields, then for the specific case 
of the matter fields. 

That D-type could have strong impact on the pattern of soft masses 
has been known for some time, particularly for the limit when the D-terms 
are small (less than the corresponding F-type contributions), 
$\ler~ \mathcal{O}(m_{3/2}^2)$ \cite{dps}.  A more
dramatic impact could be expected if the D-terms are large and as allowed 
by the cosmological constant limit, within the range, 
$ \mathcal{O}(m_{3/2}^2) \ler D \ler \mathcal{O}(m_{3/2} M_{Pl})$.
For example, considering only pure F-type breaking, leads to a typical spectrum 
of the soft masses, where the gaugino and the 
Higgsino masses are roughly proportional to the gravitino mass, $m_{3/2}$, whereas 
the scalar mass squared and the $B$-terms are proportional to $m_{3/2}^2$.
Adding large D-type sources could significantly alter this simple pattern by 
generating a splitting between the fermionic and scalar superpartners,
by an amount proportional to $D$. In the extreme limit, this would mean that the scalars
can have masses close to the intermediate scale.  Following the works of 
Ref.\cite{ibanez,ibaneztypeI}, we parameterise the soft terms 
in three particular cases (section III) : (i) mixed D and anomaly mediation 
(ii) mixed D and S mediation and (iii) mixed D and T mediation. In the mixed
D and anomaly mediated scenario, scalar masses can be everywhere
between the weak scale and an intermediate
scale whereas the gaugino masses, B-term and the $\mu$ term are proportional
to the gravitino mass, $m_{3/2}$, which can be taken close to the weak scale. 
In the mixed D and S(T) mediated scenarios, the hierarchy between scalar 
and fermionic superpartners is parameterised by an angle $\gamma_{S(T)}$, which
could be constrained by phenomenology. 

The splitting due to the D-terms could well have another important application
in understanding the origins of recently proposed ``split supersymmetry" models.
Influenced by multivacua structure in string theory as a
possible new view on the cosmological constant problem \cite{bp},
these models question the solution of the gauge hierarchy problem 
through low energy SUSY \cite{split}.  In this proposal, not all 
superpartners are required to be at a scale close to TeV. Instead, 
it is sufficient if the fermionic superpartners stay close to the 
weak scale, whereas the scalar superpartners can be present at scales as high as
$10^9$ GeV. This way, one keeps the nice features of gauge coupling
unification and the viable dark matter candidate of low energy supersymmetry,
while getting rid of unwanted features associated with large
flavour changing neutral current effects and CP violation
problems\cite{splitpheno}. In section IV, we address the question of attaining
split supersymmetry by including D-mediation. As we will see, though it is not
automatic to have exact split spectrum in these models, specifically due to the
$B$ term, we can nevertheless envisage models where it is possible to generate
hierarchical spectrum and we will present explicit models 
of this type.

So far we have not addressed the issue of the origin of such large D-terms. 
We address this issue in sections V and VI. Unlike in the
heterotic case, in Type I/II string theories, Fayet-Iliopoulos terms can
appear at the tree level and thus it is possible to generate SUSY breaking
with large D-terms. We will present an explicit example in the context of
intersecting D-branes Type I orientifold models with four stacks of D9
branes, each stack containing four coincident branes. 
 However, a related question concerns
the stabilisation of the moduli as these FI terms are field-dependent. 
We find that standard mechanisms like gaugino condensation, suitably combined
with other mechanisms of moduli stabilisation as, e.g. three-form fluxes
in IIB orientifolds, are still applicable
even in the limit of large D-terms. We present an example detailing
this point. We close with a summary. 
A preliminary version of our results was reported in \cite{dv}.

\section{General Expressions Including D-breaking}
In the following, we will present general expressions for the soft
terms including D-type supersymmetry breaking terms. As is the case
with any general analysis, we will not address the question of the
origins of these SUSY breaking \textit{vevs} for either F-terms or
D-terms. We will assume SUSY to be broken with both these types of
breaking and proceed to derive the soft terms.  As a starting
point, let us recall the form  the scalar potential
in supergravity\footnote{Most of the expressions are presented in
Planck units, namely, we set $M_P = 1$. However, at many instances,
we keep $M_P$ explicitly to make the discussions clearer.}:
\be \label{potential} V = e^G (G^M ~G_M
- 3) + {1 \over 2} \sum_A g_A^2 D_A^2 . \ee Here
$G = K + \ln|W|^2$, with $K$ being the K{\"a}hler potential and W, the
superpotential and $1/g_A^2 = Re{f_A}$, where $f_A$ is the gauge kinetic
function. The $F$ terms in the scalar potential are given by
$ G_M = \partial G/ \partial z^M $, where $z$ represents the
scalar part of a chiral superfield. The index $M$ runs over all the
chiral superfields present,  matter as well as hidden sector and/or
moduli fields. The D-terms, $D_A$ carry the obvious notation with
the index $A$ running over all the $U(1)$ factors present\footnote{Note
that the D-terms can be explicitly given in terms of the fields, derivatives
of the K{\"a}hler potential and a FI term. We will make use of this form
in a later subsection. For the present, we just note that we consider FI
terms to be moduli dependent.}.

While deriving the soft terms, a couple of constraints need to be satisfied. 
First, at the minimum, both $D$ and $F$ terms contribute to supersymmetry 
breaking and thus to the vacuum energy. This can be canceled by 
the superpotential (W) \textit{vev} which gives mass to the gravitino. 
We will impose this fine-tuning condition on the potential. This means:
\be
\label{vaccond1}
 <V> \ = \
< e^G (G^M ~G_M - 3) + {1 \over 2} \sum_A g_A^2 D_A^2 > \ = \ 0 \
. \ee Second is the necessary condition for the existence of the
minima: $\label{vaccond2} < \partial_K V>~ =~ <\nabla_K V >~ =~ 0.$ 
Here $\nabla$ denotes the covariant derivate on the K{\"a}hler manifold 
defined by $\nabla_K V_M = \partial_K V_M - \Gamma_{KM}^L V_L$. Using the 
definition of the potential, eq.(\ref{potential}) and eq.(\ref{vaccond1}), 
this implies\footnote{Strictly speaking, there is a contribution proportional 
to the derivative of the gauge kinetic function in the minimisation condition,
 eq.(\ref{vaccond2a}).  As we are concerned with the general expressions 
for the matter field soft terms, these contributions will be proportional to 
matter field \text{vev}s which are much smaller than the moduli 
\textit{vev}s and therefore we will neglect them here.}:
\be \label{vaccond2a} <e^G ( G^M~\nabla_K G_M  + G_K ) +
\sum_A g_A^2 D_A (\partial_K D_A - {1 \over 2} G_K D_A ) > = 0 \ee
We will use the eqs.(\ref{vaccond1},\ref{vaccond2a})  while
deriving general expressions for the soft terms. In the present subsection
we will not distinguish between the matter and hidden/moduli fields, but
present generic expressions for the various scalar couplings in the theory. 
To start with, we will consider the case of the scalar mass squared
matrix, which is defined as
\be
M_0^2 = \begin{array}{|cc|}
m^2_{I \bar{J}} & m^2_{I J} \\
m^2_{\bar{I} \bar{J}} & m^2_{\bar{I} J} \end{array} ~,
\ee
where the various entries are defined by:
\bea
m^2_{I \bar{J}} =~ <\partial_I \partial_{\bar{J}} V>~ =
~< \nabla_{I} \nabla_{\bar{J}} V > \\
m^2_{I J} =~ <\partial_I \partial_J V>~ =~ < \nabla_{I} \nabla_{J} V >.
\eea
Using the definition of the potential in eq.(\ref{potential}) and
the conditions, eqs.(\ref{vaccond1}, \ref{vaccond2a}), we find
the most general expressions for the bilinear couplings to be of the 
form :
\bea
\label{genscalar}
m^2_{I \bar{J}} &=& e^G ( G_{I \bar{J}} + \nabla_{I} G^{\bar{K}}
\nabla_{\bar{J}} G_{\bar{K}} - R_{I\bar{J}K\bar{L}} G^K
G^{\bar{L}} ~)
+ {1 \over 2} \sum_A g_A^2 D_A^2 ( G_{\bar{J}} G_I - G_{I \bar{J}} ) \nonumber \\
&-& \sum_A g_A^2 D_A (G_{\bar{J}} \partial_I D_A  + G_I \partial_{\bar{J}} D_A -
\partial_I \partial_{\bar{J}} D_A )
+  \sum_A g_A^2  \partial_I D_A \partial_{\bar{J}} D_A  \ , \\
\label{genbilinear} m^2_{IJ} &=& e^G(2  \nabla_J G_I + G^K
\nabla_I \nabla_J G_K ) - \sum_A g_A^2 D_A ( G_J \partial_I D_A +
G_I \partial_J D_A -\nabla_I \nabla_J D_A  ) \nonumber \\  &-& {1
\over 2} \sum_A g_A^2 D_A^2 ( G_I G_J + \nabla_I G_J + {1 \over 2}
g_A^2 \partial_I \partial_J f_A) + \sum_A
g_A^2
\partial_I D_A \partial_J D_A \ , \eea where we have neglected the
vacuum brackets for simplicity\footnote{From now on we will neglect
vacuum brackets in the rest of the paper, unless and otherwise specified.}. 
Here, $f_A$ represents the gauge kinetic function. The term containing
the second derivative $\partial_I \partial_J f_A$ gives contributions
to the $B_{\mu}$ term from operators of the form $\int d^2 \theta \ W^{\alpha}
W_{\alpha} H_1 H_2$ in superfields. A similar term of the form, 
$\partial_I f_A \partial_{\bar{J}} \bar{f}_A$ could contribute to 
$m^2_{I \bar{J}}$. However since this contribution is proportional to the 
\text{vev}s of the matter fields, as we will discuss in the next section, 
we neglect it here\footnote{For the same reason, we do not write down
the contributions from gauge kinetic function in the $A_{ijk}$ term discussed
below.}. 
The function, $R_{I\bar{J}K\bar{L}}$ represents the Riemann (curvature) tensor
of the K{\"a}hler manifold whose definition can be found in any
of the standard texts \cite{wessbagger}. The next step would be 
to derive the expression for the trilinear couplings, which we define
as\footnote{For MSSM fields this definition is equivalent to the
naive one of using ordinary derivatives giving the A-term. }: \be A_{IJK} = <
\nabla_{I} \nabla_{J} \nabla_{K} V
> \ee which takes the form : \bea \label{genaparam} A_{IJK}&=& e^G
\left( G_K (2 \nabla_J G_I + G^M \nabla_I \nabla_J G_M ) + G_J (2
\nabla_K G_I + G^M \nabla_I \nabla_K G_M ) \right. \nonumber \\
&+& \left.  G_I (2 \nabla_K G_J + G^M \nabla_J \nabla_K G_M ) + 
2 \nabla_I \nabla_K G_J + \nabla_J \nabla_K G_I + G^M \nabla_I
\nabla_J \nabla_K G_M  \right) \nonumber \\
&-& g_A^2 D_A (\nabla_I D_A -
{1 \over 2} G_I D_A)(G_J G_K + \nabla_J G_K) 
- g_A^2 D_A (\nabla_J D_A - {1 \over 2} G_J D_A)(G_I G_K + \nabla_I G_K)
\nonumber \\
&- &g_A^2 D_A (\nabla_K D_A - {1 \over 2} G_K D_A)(G_I G_J + \nabla_I G_J)
- {1 \over 2} g_A^2 D_A^2 \left( G_I G_J G_K + G_I \nabla_J G_K  \right. \nonumber \\
&+& \left. G_K \nabla_I G_J + G_J \nabla_I G_K + \nabla_I \nabla_J G_K \right)
+ g_A^2 \left( \nabla_I \nabla_J D_A \nabla_K D_A + \nabla_J D^A
\nabla_I \nabla_K D_A \right. \nonumber \\
&+&  \left. \nabla_I D_A \nabla_J \nabla_K D_A +
D_A \nabla_I \nabla_J \nabla_K D_A \right). \eea  
Note that as for the $B$ term, there can be contributions to the A-term
also from gauge kinetic function, which can be represented by operators
of type $ \int d^2 \theta W^\alpha W_\alpha h_{ijk} Q_i Q_j Q_k$ with $Q_i$ representing
the matter fields. These are typically of the order $m_{3/2}^2/M_{Pl}$ and
thus they give negligibly small contributions
unless $m_{3/2}$ has intermediate scale values.
In the next subsection, we will use these expressions to get the 
expressions of soft masses for the matter fields. 
\subsection{General Expressions of soft terms for Matter Fields}
We will define matter fields by setting their \textit{vev}s to zero. 
This would mean that both the $F$ and $D$ contributions proportional
matter field vevs to be zero at the leading order. 
Thus,  we have: 
$$\label{mattervev}
\langle \Phi^i \rangle = 0 \quad, \quad  \langle G^i \rangle = 0 \quad, \quad 
 \langle \partial_i D_A \rangle = 0 \quad,  \quad $$
with $\Phi$ representing the scalar part of a matter field. 
From now on, to distinguish matter and hidden/moduli fields, 
we denote matter (moduli/hidden sector) fields by using latin(greek) indices. 
To derive the soft terms for matter fields from the general scalar couplings 
presented in the previous section, along with using the definitions above,  
we have to remove the supersymmetric contributions from them. 
Further, we identify the gravitino mass to be $m_{3/2} = <e^{G/2}>$. 
Taking all these modifications in consideration, the final set of 
equations are of the form: 
\bea \label{mattscalarf} m^2_{i \bar{j}} &=& m_{3/2}^2~(
G_{i \bar{j}}  - R_{i\bar{j}\alpha\bar{\beta}} G^\alpha
G^{\bar{\beta}} ~) - {1 \over 2} \sum_A g_A^2 D_A^2  G_{i \bar{j}}
+ \sum_A g_A^2 D_A \partial_i \partial_{\bar{j}} D_A \ , \\
\label{mattbilinearf}
m^2_{ij} &=& m_{3/2}^2~ (2  \nabla_i G_j + G^\alpha \nabla_i \nabla_j G_\alpha )
- {1 \over 2} \sum_A g_A^2 D_A^2 (\nabla_i G_j + {g_A^2 \over 2}
 \partial_i \partial_j f_A) + \sum_A g_A^2 D_A \nabla_i
 \nabla_j D_A \ , \\
\label{mattaparamf}
 A_{ijk}&=& m_{3/2}^2 \left( 3 \nabla_i \nabla_j G_k +
G^\alpha \nabla_i \nabla_j \nabla_k G_\alpha  \right)
 - {1 \over 2} \sum_A g_A^2 D_A^2  \nabla_i \nabla_j G_k
 + \sum_A g_A^2 D_A \nabla_i  \nabla_j \nabla_k D_A \ , \\
\label{gauginof} \mu_{ij} &=& m_{3/2} \ \nabla_i G_j  \quad ,
\quad M^A_{1/2} = { 1 \over 2} ( Re f_A)^{-1} m_{3/2} f_{A \alpha}
G^\alpha \ , \eea 
where we have now also supplemented the scalar equations with those for
the $\mu$ and the gaugino masses. In the above 
$f_{A \alpha} = \partial f_A/\partial z^\alpha$. 
Note that these  expression reduces to the standard 
form\cite{wessbagger,fabio,ibanez} in the limit where $D_A$ goes
to zero. 

Note that the above soft terms are not in a
canonically normalised basis for the kinetic terms. This can be
seen from their action which has the form $g_{i \bar{j}}
\partial z^i \partial z^{\bar{j}} - m^2_{i \bar{j}} z^i
z^{\bar{j}}$, where $z^i$ represents a matter scalar field. To go
to the normalised basis, one can define vielbeins such as :
$ z^i = e^i_a z^a \;,\;\;\; z^{\bar{j}} = e^{\bar{j}}_{\bar{b}} z^{\bar{b}} $
such that $e^i_a e^{\bar{j}}_{\bar{b}} = g^{i \bar{j}}$.  Using this
transformations, we have the soft mass in the normalised basis to be given
by $$ \bar{m}^2_{a \bar{b}} = e^i_a m^2_{i \bar{j}} e^{\bar{j}}_{\bar{b}}, $$
where $\bar{m}^2_{a \bar{b}}$ represents the normalised masses.
Similar analysis can be extended for other soft terms.
In order to keep a compact notation, we however do not present
the general expressions in the normalised form.

While these expressions are given for the tree level potential,
higher order corrections can play a significant role, depending on
the specifics of the model of supersymmetry breaking. In models
with small tree-level contributions, the dominant set of
corrections are of anomaly mediated type\cite{amsb1} which are
proportional to the gravitino mass $m_{3/2}$. These contributions
are typically not modified in the presence of D-terms and have to
be anyway included. For the gauginos, the most general form of
these expressions have been presented in \cite{amsb2} and are
given by \be M_{~1/2}^{'~A} = - {g^2_A \over 16 \pi^2} \left( 3
T_G^A - T_R^A - (T_G^A - T_R^A) K_{\alpha} G^{\alpha} - {2 T_R^A
\over d_R^A} (\log \det K|_R^{"} )_{,\alpha} G^{\alpha} \right) \
m_{3/2} \ . \ee Here, $T_G$ is the Dynkin index of the adjoint
representation, normalised to $N$ for $SU(N)$, $T_R$ is the Dynkin
index associated with the representation $R$ of dimension $d_R$,
normalised to $1/2$ for the fundamental of $SU(N)$ and $K|_R^{"}$
is the K{\"a}hler metric restricted to the representation $R$. This
expression reduces to the following when all the \textit{vev}s are
much less than $M_P$: \be M_{1/2}^{'A } \ = \ - \ { g^2_A b_0^A
\over 16 \pi^2} \ m_{3/2} \ , \ee where the beta function $b_0^A$
was given as $3T_G^A - T_R^A$ in the previous expression.
In addition to the gauginos, the scalar mass terms as well as the
B-term  and the A-terms receive corrections. In the case of gauginos,
as long as the tree-level F-term  contributions are present, the anomaly
mediated contributions remain sub-dominant, whereas in the case of
scalar soft terms, both the D-term as well as the F-term contributions
have to be suppressed for the anomaly mediated contributions to dominate.
\subsection{Implications of large D-terms on the soft parameters}

Eqs.(\ref{mattscalarf}-\ref{gauginof}) give the modified expressions for
the soft terms after including non-zero D-type SUSY breaking contributions
in supergravity. Whereas the scalar couplings receive corrections
from the D-type terms, the gaugino masses
are unaffected by D-mediated effects. The $\mu$ 
term, could be visualised as a soft mass in supergravity by using the Giudice-Masiero
mechanism\cite{gm}. The expression presented in the previous sub-section takes care
of this situation and it is seen that $D$ terms do not effect the $\mu$ term either. 
However, the exact implications on the soft terms by the inclusion of the D-terms
depend on (a) the structure of the D-terms and (b) the magnitude of them. 
We will address these two issues below.  

In the presence of anomalous non-linearly realised abelian gauge symmetries
\bea
&& \delta V_A \ = \ \Lambda_A + {\bar \Lambda}_A \ , \ \delta z^i \ = 
\ \Lambda_A
X^i_A  z^i  \ \equiv \mathcal{V}_A^i \Lambda_A \ , \nonumber \\ 
&& \delta T^{\alpha} \ = \ \eta_A^{\alpha} \Lambda_A \equiv  \mathcal{V}_A^{\alpha} 
\Lambda_A \ , \label{s01} 
\eea 
where $\mathcal{V}_A^I$ are the Killing potentials, the auxiliary D-terms, 
defined by
\be
\partial_{\bar J} D_A \ = \ \mathcal{V}_A^I K_{{\bar J},I} 
\ = \ \mathcal{V}_A^i K_{{\bar J},i} +
\mathcal{V}_A^{\alpha} K_{{\bar J},\alpha} \label{s02}
\ee
 are explicitly given by
\be
\label{Dtermdef}
D_A \ = \ z^I X_{I}^A {\partial K \over \partial z^I} ~+~ \xi_A \ = \ 
 {\bar z}^{\bar I} X_{I}^A {\partial K \over \partial {\bar z}^{\bar
I}}   ~+~ \xi_A \ , \ \xi_A \equiv \eta^\alpha_A \ \partial_\alpha K 
\ , \label{s1}
\ee
where $X^A_I$ represents the $U(1)_A$ charges of the fields $z^I$
and $\xi_A$ denotes the Fayet-Iliopoulos term for the $U(1)_A$ factors.
Note that the equality
between the two last terms is a straightforward consequence of the gauge
invariance of the K{\"a}hler potential. We consider the Fayet-Iliopoulos
terms to be moduli dependent and we will not explicitly discuss here the various
possible mechanisms of moduli 
stabilisation\footnote{After moduli stabilisation, the anomalous
$U(1)$'s become gauged R-symmetries \cite{freedman}.}.
We have in the vacuum, after setting the matter fields \textit{vevs} to zero
\bea
&& \langle \partial_j D_A \rangle =
\langle \bar{v}_{\bar \beta} X^A_{\bar \beta} K_{\bar{j} \beta} +
  \eta_A^{\bar{\alpha}} K_{j \bar{\alpha}} \rangle = 0\quad , \quad
\langle \nabla_i \nabla_j D_A \rangle = 0 \ , \nonumber \\
 \langle \partial_i \partial_{\bar j} D_A \rangle &=& K_{i {\bar j}} X_{i}^A
+ (\bar{v}^{\bar{l}} X^{\bar{l}}_A \partial_{\bar{l}} + \eta^{\bar{\alpha}}_A
\partial_{\bar{\alpha}} )~K_{i \bar{j}} \;\; , \;
\langle \nabla_i \nabla_j  \nabla_l D_A \rangle = 0
 \label{s2} \eea
By using (\ref{s2}), the soft terms for the matter fields reduce to 
\bea
\label{mattscalarD} m^2_{i \bar{j}} &=& m_{3/2}^2 \left( G_{i \bar{j}}-
R_{i\bar{j}\alpha\bar{\beta}} G^\alpha G^{\bar{\beta}} \right)
+ \sum_A  g_A^2 D_A ~\left( X^A_i 
+  \bar{v}_{\bar{l}} X_{\bar{l}}^A \partial_{\bar{l}}
+  \eta^{\bar{\alpha}}_A \partial_{\bar{\alpha}} -
{1 \over 2}  D_A \right)G_{i \bar{j}} \ , \\
\label{mattbilinearD}
m^2_{ij} &=&  m_{3/2}^2~ \left( 2 \nabla_i G_j +~G^\alpha \nabla_i \nabla_j
G_\alpha \right) - {1 \over 2} \sum_A g_A^2 D_A^2
( \nabla_i G_j + {g_A^2 \over 2}  \partial_i \partial_j f_A)
 \ , \\
\label{mattaparamD}
 A_{ijk}&=& m_{3/2}^2 \left( 3 \nabla_i \nabla_j G_k + G^\alpha \nabla_i
   \nabla_j \nabla_k G_\alpha \right) - {1 \over 2} \nabla_i \nabla_j G_k
\sum_A g_A^2 D_A^2 \ .
\eea

Let us now try to quantify how large the D-terms can be. To do
this, let us consider the following generic forms for the K{\"a}hler
and the superpotential : \bea \label{genkahler} K &=&
\tilde{K}(T_\alpha, T_{\bar{\beta}} ) + H_{i \bar{j}}(T_{\alpha},
T_{\beta}) Q_i Q^\dagger_{\bar{j}} +
(Z_{ij}(T_{\alpha},T_{\bar{\beta}}) Q_i Q_j
+ h.c) + \ldots \\
\label{gensuper} W&=& Y_{ijk}(T_{\alpha}) Q_i Q_j Q_k +
\tilde{W}(T_{\alpha}) + \ldots, \eea where $T_{\alpha}$ represent
moduli/hidden sector fields and $Q_i$ represent the matter fields.
Using these equations let us now revisit the condition
(\ref{vaccond1}) \be \label{vaccond1expa} m_{3/2}^2 \left(
K^{\alpha\bar{\beta}} (K_\alpha K_{\bar{\beta}} + {M_P^2 \over W}
(K_{\alpha} W_{\bar{\beta}} + W_{\alpha} K_{\bar{\beta}}) + {M_P^4
\over |W|^2} W_{\alpha} W_{\bar{\beta}}) -3 M_P^2 \right)
+ {1 \over 2}~ g_A^2 D_A^2 = 0 \ . \\
\ee From the above we see that, as long as the D-terms are in the
limit, $D ~\sim~ \mathcal{O}(m_{3/2}^2)$, they would not
contribute significantly to the vacuum energy. However, when they
lie within the limit \be \label{dtermlimit} m_{3/2}^2 \ler D_A
\ler m_{3/2} M_P, \ee they could be contributing significantly.
The upper limit is obtained when one assumes D-term contributions
to dominate over the F-term contributions or are of the same order
as them. This particular limit is what we are interested in the
present work as this has not been exploited in a general manner as
presented here. From the generic set of soft parameters presented
above, it is obvious that splitting between fermionic and scalar
superpartners can be `naturally' achieved once the D-terms lie
within the above range. Quantitatively, if in a given model the
gravitino mass is of $\mathcal{O}$(1 TeV), the upper limit on the
D-term would be of the order of intermediate scale $\sim$
$(10^{10})$ GeV. It is obvious that as one increases the gravitino
mass closer to the intermediate scale $\sim~ (10^9-10^{12})$ GeV,
the upper bound on the D-terms become close to the GUT scale.
These upper bounds are essentially the magnitude required to
cancel the cosmological constant in the limit where the F-terms
tend to zero.

Given this limit, let us now try to understand in more detail how large
D-terms would generate large splittings between superpartners.
The equations for the gaugino and $\mu$-term remain unchanged as we
have mentioned. The following features of the spectra are easy to
extract without actually being specific about the model:
\begin{itemize}
\item (i). \textit{Scalar Mass Terms}:
The most dominant contribution to the scalar masses from the 
$D$-terms are the ones which are linear in $D$  which for
 $m_{3/2} \sim$ TeV push the scalar masses to intermediate energy 
scale. Note that these terms depend on the charges
of the fields under the additional $U(1)$ gauge group, thus
putting a constraint that these charges to be of definite sign. If all the
three generations of the sfermions have the same charges under the
$U(1)$ groups, this term would also be universal.  Otherwise, there 
are off-diagonal entries which are generated in the mass
matrices, which could of suppressed by some powers in the
expansion parameter $\epsilon_{\beta} = v_{\beta}/M_P$, with $v_{\beta}$
representing the \textit{vev} of some flavon field. 

\item (ii). \textit{Higgs mass terms and the $B_\mu$}: The Higgs masses
follow almost the same requirements as the soft masses. Usually,
their charges are linked with the Giudice-Masiero mechanism\cite{gm}.
The $B_\mu$ term is however special. Unlike the Higgs mass terms,
it does not receive large contributions from D-terms, whose contributions
can be utmost of $\mathcal{O}(m_{3/2}^2)$. If the splitting between
the Higgs masses and the $B_\mu$ is too large, it could lead to unphysical
regions in $\tan \beta$. This could be easily seen by noting that 
\be
\label{ewsb}
\sin 2 \beta = {2 B_\mu \over m_{H_1}^2 + m_{H_2}^2 + 2 \mu^2}.
\ee
In the limit of large Higgs mass parameters $m_{H_1}^2,~ m_{H_2}^2$, 
one has to think of ways to enhance
the $B_\mu$ term. We will present one such example in the next section. 

\item (iii). \textit{A-terms}: Even if the D-terms are large, the
A-terms are typically proportional to $\mathcal{O}(m_{3/2})$. No
large enhancement is present. This is expected as A-terms break
R-symmetries. They get related to the D-terms due to the constraints
of cosmological constant cancellation, but as the scale of R-symmetry
breaking is set by the gravitino mass, this naturally sets the A-terms
to be of same order.

\item(iv). \textit{Gaugino Masses}: All along we have been commenting 
that the presence of SUSY breaking D-terms would not change the results 
for the gaugino masses presented there. This is only true as long as there 
are no additional fermions in the model. In the presence of additional 
fermions and non-zero
D-terms, gauginos can get Dirac masses through operators of the
form \cite{nelson}
\be
h_a \int d^2 \theta \ { \chi^a W_\alpha^a  W_X \over M_Pl } \ = \ h_a {\langle D_X \rangle \over M_P} \psi^a  \lambda^a \ + \cdots \ = \
m_D^a \psi^a  \lambda^a + \cdots \ ,
\ee
where $\chi^a$ represent here fields in the adjoint representation of
the Standard Model gauge group with (mirror fermions) which mix with the
gauginos and $m_D^a$ represent the Dirac mass for the gauginos. These
mixing terms could lead to the Majorana masses for the gauginos by a
seesaw mechanism $\sim~(m_D^{a})^2/M_a $ if the mirror fermions obtain 
large R-symmetry breaking Majorana masses, $M_a$. In the present work, 
we do not concentrate on building models of this type.
\end{itemize}
\section{Parametrization of soft terms in Type I/II string models with 
large D-terms}
\noindent The soft terms in effective string supergravities from
Type-I/II string theories have been parameterized in
\cite{ibaneztypeI} where pure $F$-type breaking has been assumed.
In the present section we will extend this analysis by considering
D-type SUSY breaking terms too. In each of this case, we present
parameterizations of the soft terms which could be readily be
useful for phenomenological studies.
\subsection{D-dominated  supersymmetry breaking}
The first case we consider is that of a scenario where F-terms are
absent or negligible. We assume that supersymmetry breaking is
achieved by pure D-terms. However, we will still require that the
gravitino get a mass. This would enable us to cancel the
cosmological constant even in the pure D-breaking
limit\footnote{An earlier proposal in this direction has been
presented in \cite{dvalipomarol}.}. The scale of the gravitino
mass is assumed to be not very far from the weak scale.  With these
conditions, the potential, eq.(\ref{potential}) takes the form:
\be V = {1 \over 2} \sum_A g_A^2 D^2_A - 3 m_{3/2}^2 M_P^2 \ . \ee
It is obvious from the above equation that requiring that the
potential should vanish at the minimum (for the cosmological
constant cancellation), implies that the D-terms should be \be
<D> = { \sqrt{6} \over g} \ m_{3/2} M_P \ . \ee
A more subtler constraint comes from the existence of a minimum,
eq.(\ref{vaccond2a}). In this limit, it takes the form $g_A^2 D_A
(\partial_{\beta} D_A) = 0$. It is clear that for a single $U(1)$
gauge group, this would mean at the minimum either the
\textit{vev} to vanish or the D-term to vanish. Both these
conditions are not acceptable to us. The situation would not change
even if one adds more flavon fields. Thus we rule out the case of
single $U(1)$ with pure D-breaking. The minimum case we can think 
of is that case with two $U(1)$ gauge groups with two charged fields.

We parameterise the SUSY breaking D-terms, consistently with 
the vanishing of the cosmological constant, 
as \be <D_A> = { \sqrt{6} \over g_A}  \
\theta_A \ m_{3/2} M_P \ , \ee where $\theta_A$ are defined such
that $\sum_A \theta_A^2 = 1$. Then the soft terms reduce to the
following form : \bea m^2_{i \bar{j}} &=&  - 2 m_{3/2}^2 \ G_{i
\bar{j}}+ \sqrt{6} \ m_{3/2} M_P \ \sum_A g_A \theta_A 
( X^A_i +{v}_{\bar{l}} X_{\bar{l}}^A \partial_{\bar{l}}
+  \eta^{\bar{\alpha}}_A \partial_{\bar{\alpha}} )G_{i \bar{j}}
\nonumber \ , \\
m^2_{ij} &=& - m_{3/2}^2 \ \left( \nabla_i G_{j} + {3 \over 2} \sum_A
  g_A^2 \theta_A^2 \partial_i \partial_j f_A \right) \ , \nonumber \\
\mu_{ij} &=& m_{3/2} \ \nabla_i G_j  \ , \nonumber \\
A'_{ijk} &=& m_{3/2} \lambda_{ijk}(\gamma_i + \gamma_j + \gamma_k) \ ,
\nonumber \\
M_{1/2}^{'~A } &=& - \ { g^2_A b_0^A \over 16 \pi^2} \ m_{3/2} \ .
\label{d1} \eea The gaugino masses vanish at the tree level in
this limit. They are generated by anomaly mediated contributions
as listed above. Similar thing happens for the A-parameters, which
are determined by their anomalous dimensions ($\gamma_i$) as given
above. Note that the above mass formulae are given at
the high scale. One has to evolve these masses at the weak scale
to make contact with weak scale phenomenology.  The present
scenario describes a new situation where the non-holomorphic
scalar soft masses are given by dominant D-type supersymmetry
breaking terms, whereas the gauginos, described by the
beta-functions, the supersymmetric fermion masses
(in particular the $\mu$ term of MSSM) are proportional to the
gravitino mass and have therefore much lower values.
If all the $U(1)$ groups are in the visible sector with 
large D-terms and positive charges, such a situation is not phenomenologically
viable, since there is no possibility of tuning one Higgs doublet to
be very light.  However, if some of the $U(1)$ lie in the hidden sector with 
some others in the visible sector and the angles $\theta_A$ in 
the visible sector are all small, then the scenario with pure D-breaking becomes viable. 
In this last case, all the soft terms can be at the TeV scale, thus making contact with a low
energy physics of the MSSM type.  It would be
interesting to see how this new structure of soft terms would feature
with respect to low-energy constraints like electroweak symmetry
breaking, dark matter, LEP Higgs bounds and other constraints.
Note that a situation like split SUSY could be difficult to incorporate
here. 
\subsection{D-breaking with dilaton and moduli supersymmetry breaking }
The above discussion presents an extreme situation \textit{i.e.}
completely absent F-type breaking. However such an extreme limit
is not required to realise split supersymmetry breaking.  The
general analysis presented in the previous section shows that it
is enough to have $g_A^2 D_A >> m_{3/2}^2$. We present here soft
terms for a case where, for simplicity, there is only one $U(1)$
large D-term and we assume that the auxiliary field of the dilaton
or the overall modulus superfields also contribute to
supersymmetry breaking.

Note that such a situation can arise naturally when one considers
effective lagrangians of Type I string theory for an orientifold
with only $D9$ branes. We provide expressions for the case of
orbifold theories (Calabi-Yau spaces are also particular cases of
the expressions below) in the large volume limit. In this limit,
the gauge kinetic function and the K{\"a}hler potential $K$ will
have the general form \cite{quevedoreview}: \bea f_A^B &=& S \
\delta_A^B + \ldots\ , \nonumber \\  K &=& -\log ( S + S^\dagger) - 3 \log
( T + T^\dagger - \delta_{GS} V)  + \sum_i (T + T^\dagger -
\delta_{GS} V)^{n_i}|\phi_i|^2 \nonumber \\
&+& \sum_{ijk} \left( Z_{ijk }(T + T^\dagger - \delta_{GS}
V)^{n_i} {\bar \phi_i} \phi_j \phi_k + {\rm h.c.} \right) + \cdots
\ , \label{d01} \eea where we have used the by now standard
notation with $S$ representing the dilaton field, $T$ representing
the overall volume modulus, $\phi_i$ represent matter fields and
$n_i$ modular weights of matter fields. We are assuming from now
on that the modulus $T$ is the one mixing with the anomalous
$U(1)$ gauge field, such that the gauge invariant combination $T +
T^\dagger - \delta_{GS} V$ should consistently appear in the
K{\"a}hler potential and in the couplings to the matter fields. This
can be explicitly realized in intersecting brane models, as we
will illustrate later on. The last term in the K{\"a}hler potential in
(\ref{d01}) accommodate the possibility of $\mu$ terms and
simultaneously, that of the $B_\mu$ term. We parameterize the
supersymmetry breaking contributions from the two sets of
auxiliary fields as : \be <G_S> = \sqrt{3} \ ({M_P \over S +
S^\dagger}) \ \cos \gamma_S \quad , \quad <D> = {\sqrt{6} \over g}
m_{3/2} M_P \sin \gamma_S \ . \ee We then obtain the soft terms

\bea m^2_{i\bar{j}} &=& (1 - 3 \sin^2 \gamma_S) m_{3/2}^2
G_{i\bar{j}} + \sqrt{6} \  g \ m_{3/2} M_P \sin \gamma_S (X_i +
\bar{v}_{\bar{l}} X_{\bar{l}} \partial_{\bar{l}}
+  \delta_{GS} \partial_{\bar{T}} ) G_{i \bar{j}} \nonumber \\
m_{ij}^2 &=& ( 2 - 3 \sin^2 \gamma_S ) m_{3/2}^2 \nabla_i G_j
\ - {3 \over 2} m_{3/2}^2 g^2 \sin^2 \gamma_S \ \partial_i \partial_j
f \ , \nonumber \\
 A_{ijk}&=& 3 m_{3/2}^2 \cos^2 \gamma_S \nabla_i \nabla_j G_k \ , \nonumber \\
M_{1/2}^A &=&  {\sqrt{3} \over 2}  m_{3/2} \cos \gamma_S \ , \eea
whereas the $\mu$ term is unchanged (\ref{d1}). In the
complementary case where the only F-type source of supersymmetry
breaking comes from the $T$ field, the appropriate parametrization
is \be <G_T> = - 3 \ ({ M_P \over T + T^\dagger}) \ \cos \gamma_T
\quad , \quad <D> = {\sqrt{6} \over g} m_{3/2} M_P \ \sin \gamma_T
\ . \ee The soft terms in this case are given by \bea
m^2_{i\bar{j}} &=& (1 +n_i \cos^2 \gamma_T - 3 \sin^2 \gamma_T  )
m_{3/2}^2 G_{i\bar{j}} + \sqrt{6} \  g \ m_{3/2} M_P \sin \gamma_T
(X_i + \bar{v}_{\bar{l}} X_{\bar{l}} \partial_{\bar{l}}
+  \delta_{GS} \partial_{\bar{T}} ) 
 G_{i\bar{j}} , \nonumber \\
m_{ij}^2 &=& [ 2 + (n_i+n_j) \cos \gamma_T - 3 \sin^2 \gamma_T ]
m_{3/2}^2 \nabla_i G_j - {3 \over 2} m_{3/2}^2 g^2 \sin^2 \gamma_T 
\ \partial_i \partial_j f \ , \nonumber \\
  A_{ijk}&=& m_{3/2}^2 \ [ 3  \cos^2 \gamma_T + (n_i+n_j+n_k) \cos \gamma_T ]
\ \nabla_i \nabla_j G_k \ , \nonumber \\
M_{~1/2}^{'~A} &=& - {g^2_A \over 8 \pi^2} \left( 3 T_G^A \sin^2
{\gamma_T \over 2} - T_R^A (\sin^2 {\gamma_T \over 2} - (1+n_i)
\cos \gamma_T) \right) \ m_{3/2} \ . \label{d2}\eea Several
simplifying assumptions were used in deriving (\ref{d2}). For
reasons already explained, the analytic scalar masses come from a
Giudice-Masiero term in the K{\"a}hler potential of the type
$\phi^{\dagger} Q_i Q_j + {\rm h.c.}$, where $\phi$ is a flavon
type field with a large vev. The Yukawa couplings were assumed, in
the large volume limit, to become T-modulus independent, otherwise
new contributions appear in the trilinear A-terms. The natural
values of modular weights for charged D9 branes charged fields are
$n_i = -1$.
Finally for phenomenological studies, the angles $\gamma_{S,T}$
can be used as independent parameters to be constrained by low energy
physics.

\section{Split Supersymmetry}
The requirement of split supersymmetry type soft spectra are as
follows : \\
\noindent (i) Scalar soft terms :
$m_{\tilde{f}}^2~ \sim ~ \mathcal{O}(10^6 - 10^{15})$ GeV,
$(\tilde{f} ~= ~Q, ~u^c,~d^c, ~L,~e^c)$ \\
\noindent (ii). Higgs mass parameters
$m_{H_1}^2~ \sim ~m_{H_2}^2 \sim~ B_\mu ~\sim  ~
\mathcal{O}(10^6 - 10^{15})$ GeV, with one of the Higgs mass eigenvalues
 fine tuned to be
around the electroweak scale.\\
\noindent
(iii). The gaugino masses and the $\mu$ term are around the weak scale.\\
\noindent

As a starting point, let us consider for a moment that all D-term
contributions are negligible or zero. In such a case, we see that
most likely the mass squared terms are proportional to $m_{3/2}^2$
whereas the gaugino masses are proportional to $m_{3/2}$. Thus, it is
difficult to expect a large splitting within the masses of the
superpartners in supergravity theories with pure or dominant F-type
SUSY breaking. In principle, such a splitting can be arranged by choosing
suitable parameter space within the goldstino directions in certain
classes of effective lagrangians coming from heterotic strings.
However, it is not clear how much these parameter spaces would remain
stable under radiative corrections. Another approach for creating a split
would be to assume some R-symmetries\footnote{Or even a charge symmetry
accompanied by F-breaking of charged chiral superfield, \cite{wells}.}
protecting the fermion superpartners. In this case, the gravitino mass
needs to be pushed to very high values, whereas the gauginos need another
mechanism to achieve masses close to the weak scale\cite{split,ad}.
However in this case, one has to invent a mechanism to suppress the
anomaly mediated contributions, which could involve for example no-scale
type models.

In the presence of D-terms,  it is generically difficult to realise
 split supersymmetry like models \footnote{See also \cite{kn}.}. 
From the discussion in the previous section, it was obvious that it is
just not sufficient to choose the $U(1)$ charges of the scalars to be
positive to realise the split spectrum since $B_\mu$ term does not
have large D-term contributions, we need to disentangle the $\mu$ and the
$B_{\mu}$ term by introducing a new field $X$ and allowing a term of the
type $X H_1 H_2$ in the superpotential. 
In a simple example, the field content is as
follows. The model contains an additional $U(1)$ group, with two additional
fields $X$ and $\phi$ with charges $+2$ and $-1$. The $\phi$ field
can act as a flavon field attaining a large vev close to the fundamental
scale. The superpotential and the relevant term in the K{\"a}hler potential are
 specified as
\bea
&& W = W_{SSM} + \lambda_1 X H_1 H_2 + \lambda_2 X \phi^2 + \cdots \ ,
\nonumber \\
&& K \supset \sum_i |\phi_i|^2  + (\phi^{\dagger})^2 H_1 H_2 + \cdots \ .
\eea
The scalar potential at the global SUSY level is given by
\be
\label{toymodel}
V = \lambda_2^2 (|\phi|^4 + 4 | X|^2 | \phi|^2) +
{ 1 \over 2} g^2 \ (2 | X|^2 - | \phi|^2 + \xi)^2 + \ldots \ .
\ee
For $\xi >0$, the stable extremum of the above potential and the auxiliary fields
are given by:
\bea
&&\langle \phi \rangle = {g^2 \over 2 \lambda_2^2 + g^2 } \xi \ , \
\langle X \rangle = 0 , \nonumber \\
&&\langle F_{\phi} \rangle =0 \ , \ \langle F_X \rangle =
{\lambda_2 g^2 \over 2 \lambda_2^2 + g^2 } \xi \ , \
\langle D \rangle = {2 \lambda_2^2 \over 2 \lambda_2^2 + g^2 } \xi \ .
\eea
From the above it is clear that $F_X \sim g^2 D$ and moreover of the order
of the FI term $\xi$. This is sufficient to enable the $B$ term to receive
large contributions through the term $G^X \nabla_{H_1} \nabla_{H_2} G_X$
in the eq.(\ref{mattbilinearD}). As long as $\xi$ is close to an intermediate
scale value, this model seems to replicate the split spectrum, if one fixes the
gravitino mass around 1 TeV. However,
in typical string models, the
FI term is of the $\mathcal{O}(M_{Pl}^2/16\pi^2)$ which would give a too
large contribution to the vacuum energy. One way to get the correct order 
of magnitude is by incorporating the above model into a higher dimensional theory. 
For illustration lets us consider a 5D theory compactified over $S^1/Z_2$. The
Standard Model and the $X,~\phi$ fields live on a 3D brane, whereas the
gauge fields of the $U(1)$ are allowed to propagate in the bulk. We will
use Scherk-Schwarz mechanism to break supersymmetry.  The R-symmetry is 
also broken by this mechanism giving rise to the gravitino mass.

The various scales in the problem are
$R = t M_5^{-1} \ , \ R M_5^3 = M_P^2 $, where
$t \equiv Re \ T$, the modulus field.
After canonically normalizing the various fields by
${\hat \phi}_i = \sqrt{t/3} \ \phi_i$ and at the global supersymmetry level,
the potential retains the form (\ref{toymodel}) with
 $\xi \sim M_5^2 = M_P^2/t $.
The four dimensional $U(1)$ gauge coupling is given by
$g^2 = 1/t = 1/(R M_5)$, whereas the gravitino mass is given by
$m_{3/2} = { \omega / R}$, where $\omega$ is a number of order
one. The D-term contribution to the vacuum
energy is then of the form
\be
\langle V_D \rangle \sim g^2 M_5^4 \sim m_{3/2}^2 M_{P}^2 \ ,
\ee
in the right order as required by the cancellation of the vacuum
energy in supergravity and realisation of the split spectrum. If the no-scale
structure is broken by the dynamics, the gauginos attain
their masses through anomaly mediation and thus we choose the gravitino
mass to be of the order of 100 TeV. The $\mu$ is
generated by the Giudice-Masiero mechanism and is 
$\mu \sim (<\phi>/M_5)^2 m_{3/2}$.
So, this model replicates the spectrum of the split
supersymmetry at the weak scale using large D-terms of the intermediate
scale and a 100 TeV massive gravitino.

In the light of above discussion, an important question is in
which sense the light Higgs mass tuning is preferred over the
tuning of another scalar mass. Tuning of squarks or slepton masses
is best described in terms of alignment in the $3 \times 3$ flavor
space. If sfermion mass matrices are very close to the diagonal,
i.e. off-diagonal terms are very small compared to the diagonal
ones, the tuning of a small mass eigenvalue is impossible, whereas
the tuning becomes more and more likely for off-diagonal terms of
the same order as the diagonal ones. In flavor models with a low
energy supersymmetric spectrum, the alignment of the quark-squark
and lepton-slepton mass matrices was necessary to avoid too large
FCNC effects, but a serious tension between alignment and
hierarchy of fermion masses was present, at least for models with
only one $U(1)$ factor. It is ironical that, in the limit of
evading FCNC effects by decoupling the undesirable scalar
particles, the alignment has still to be invoked in order to
minimize the likelihood of the fine-tuning of squark and  slepton
masses compared with the tuning of the light Higgs mass.

\section{Nonperturbative moduli stabilisation and large D-terms}

In string theory, the FI terms are field (moduli)
dependent. If no additional dynamics is present, the moduli fields
will always exhibit a runaway behaviour and the FI terms
disappear. We revisit here the issue of moduli stabilisation with
realisation of large D-term contributions in a context similar to, 
but having some new
features compared to the one discussed some time ago in \cite{bd}. 
As will become transparent, our analysis is also relevant 
for the issue of the uplift
of the energy density in the context of KKLT type moduli
stabilisation \cite{kklt,bkq}. The gauge group consists of the Standard Model
supplemented by a confining hidden sector group and an anomalous
$U(1)_X$. 
We consider the case of a supersymmetric
$SU(N_c)$ gauge group with $N_f$ quark flavors $Q^a_i$ and anti-quark
${\tilde Q}^a_{\bar i}$ where $a=1 \cdots N_c$ is an index in the fundamental
representation of the $SU(N_c)$ gauge group and $i,{\bar i} = 1 \cdots N_f$
are flavor indices. In the intersecting string realisation, discussed in
some detail in the next section, the hidden sector
consists of a stack of $N_c$ magnetised D9 branes in the type I
string with kinetic function $ f = S + k T$, where S is the
dilaton (super)field, T a volume (K{\"a}hler) modulus and $k$ is a
positive or negative integer determined by the magnetic fluxes in
two compact torii. The low energy dynamics
is described by $M^i_{\bar j} = Q^{a,i} {\tilde Q}^a_{\bar j}$, 
the composite "mesons" fields. In the following we denote by
$q$ $({\bar q})$ the $U(1)_X$ charges of the hidden sector quarks (antiquarks).
Since the FI terms are
T-modulus dependent, T will shift under gauge transformations
\bea
&& V_X \ \rightarrow \ V_X +  \Lambda_X  + {\bar \Lambda}_X \ , \ 
M^i_{\bar j} \ \rightarrow \ e^{-2 (q+{\bar q})\Lambda_X } \ M^i_{\bar j}
\ , \nonumber \\
&&T \ \rightarrow \ T + \ \delta_{GS} \ \Lambda_X \ , \label{m01}
\eea
where
\be
\delta_{GS} \ = \ {C_{N_c} \over k} \ , \ C_{N_c} \ = \ {1 \over 4 \pi^2}
{N_f (q + {\bar q})} \ , \label{m02}
\ee 
is uniquely fixed by the requirement that the mixed $U(1)_X SU(N_c)^2$ 
anomaly, denoted $C_{N_c}$ in (\ref{m02}),  to be exactly
canceled by the nonlinear transformation of $Im T$.  Notice that the
nonlinear transformation of $T$ forces a chiral nature of the hidden sector
with respect to the anomalous abelian gauge group, which in turn triggers
supersymmetry breaking \cite{bd}. In order to be able to write gauge invariant
mass terms for the mesons, a field with charge opposite in sign to the ones
of the mesons has to be introduced, called $\phi$ in what follows, of
charge $-1$ in our conventions.
The dynamical scale of the hidden sector gauge
group, the effective superpotential \cite{ads} and the K{\"a}hler potential are 
\bea && \Lambda = M_P \ e^{- 8 \pi^2 (S+k T)/(3N_c-N_f)} \ , \nonumber \\ 
&& W = W_0 (S) + (N_c-N_f) 
\left({\Lambda^{3N_c-N_f} \over det M}\right)^{1 \over N_c-N_f} 
+  m_i^{\bar j} ({\phi \over M_P})^{(q+{\bar q})} M^i_{\bar j} \
, \nonumber \\
&& K = - \ln \ (S+ {\bar S}) - 3 \ln \ [ T + {\bar T} - 
2 Tr (M^{\dagger} M)^{1/2} - |\phi|^2 - \delta_{GS} V] \ , 
 \label{m03} \eea 
where $ m_i^{\bar j}$ are mass parameters. 
Notice first of all that the dynamical superpotential
\be
W_{np} \ = \ (N_c-N_f) \ \left({e^{- 8 \pi^2 (S + k T)} \over 
det M}\right)^{1 \over N_c-N_f}
\ , \label{m04}
\ee
is precisely gauge invariant when the anomaly cancellation conditions
(\ref{m01})-(\ref{m02}) are satisfied. 
In order to stabilise the modulus S we invoke the
three-form NS-NS and RR fluxes. $W_0$ depends on
the modulus $S$, $S =S_0$ and eventually other (complex structure) 
moduli of the theory and stabilises them  by giving them a very large
mass. If the other relevant mass scales, the FI term and the
dynamical scale $\Lambda$ have much lower values, we can safely
integrate out these fields, by keeping the T modulus in the low
energy dynamics. The resulting lagrangian is similar to the one
invoked in the KKLT moduli stabilisation \cite{kklt} with 
a D-term uplifting of the vacuum energy \cite{bkq}. Notice however
that the simple nonperturbative superpotential 
$e^{- a T}$ considered in \cite{bkq} cannot be
gauge invariant due to the gauge transformation of $T$ and therefore,
precisely as in the heterotic case discussed in \cite{bd}, charged
hidden sector matter with appropriate charges is crucial to
define a consistent gauge invariant model. 

Minimisation with respect to $T$ in (\ref{m03}) stabilises also
the K{\"a}hler modulus. 
For notational simplicity we discuss in some detail the case of an 
supersymmetric
hidden sector $SU(2)$ gauge group with one quark flavor $Q^a$ and anti-quark
${\tilde Q}^a$ where $a=1,2$ is an index in the fundamental
representation of the gauge group. Due to the anomalous nature 
of the $U(1)_X$,
the sum of the quark and antiquark charges, equal to the $M$ meson charge,
is different from zero and, in our example, equal to $+1$.
$\phi$ is  a field of charge $-1$ which  participate
in the Yukawa coupling $\lambda \phi Q^a {\tilde Q}^a$, which
plays the role of meson mass after the spontaneous symmetry
breaking of the $U(1)_X$. The fact that the meson masses come from
a perturbative trilinear Yukawa coupling in this case is instrumental
in producing a large D-term contribution to supersymmetry breaking.   
 In order to provide explicitly the scalar potential, we define the 
canonical field $M \equiv \chi^2/2$. Then the
supergravity scalar potential can be  found to be \bea V_F&=& {1
\over r^3} \left\{ {r^2 \over 3} |\partial_T W - {3 \over r} W|^2
+  {r \over 3} \sum_{i=1}^2 |
\partial_i W + \bar{\phi_i} \partial_T
W|^2 - 3 |W|^2 \right\} \ , \nonumber \\
V_D&=& {1 \over S + \bar{S} + k'(T + \bar{T})}  \left( {3 \over r}
X_i |\phi_i|^2 + 3 {\delta_{GS}
 M_P^2 \over T + \bar{T} } \right)^2 \ , \label{ss3}
\eea where $\phi_i = \chi , \phi$, we have introduced $r
\equiv ( T + \bar{T} - \sum_i|\phi_i|^2)$, $\delta_{GS}$ represent
the Green-Schwarz coefficient of the $U(1)_X$, and $k'$ is the magnetic flux 
on the brane providing the anomalous $U(1)_X$. 

By inserting (\ref{m03}) into (\ref{ss3}) we find a model with all
moduli stabilised. If we would ignore the $U(1)_X$ dynamics, for example,
$T$ would be stabilised as in \cite{kklt} by solving $D_T W = 0$. In our
case, the minimum $T_0 = \langle T \rangle$ will be shifted due to the
D and new F contributions.
A full supergravity analysis of the vacuum
of (\ref{ss3}) is possible but cumbersome. Due to this after stabilizing
$S = S_0$ and $T =T_0$ by solving their equations of motion, we analyse the
stabilisation of the other fields, for simplicity
 at the global supersymmetry level, as in
\cite{bd}, by a suitable rescaling of the fields and Yukawa coupling
$\lambda$. For general $N_c$ , $N_f$ at the global level, the 
auxiliary fields and the scalar potential are
\bea
&& (F^{\bar M})^{\bar i}_i \ = \ 2 [(M^{\dagger} M)^{1/2}]^{\bar i}_{\bar j}
\left[ - (M^{-1})^{\bar j}_i 
\left({\Lambda^{3N_c-N_f} \over det M}\right)^{1 \over N_c-N_f}  
+  m_i^{\bar j} ({\phi \over M_P})^{(q+{\bar q})} \right] \ , \nonumber \\
&& {\bar F}_{\bar \phi} \ = \ {q+{\bar q} \over M_P}  
({\phi \over M_P})^{q+{\bar q}-1} Tr (m M) \ , \nonumber \\
&& D_X \ = \ (q+{\bar q}) Tr (M^{\dagger} M)^{1/2} - |\phi|^2 +
k \mu^2 \ , \nonumber \\
&& V \ = \  |F_{\phi}|^2 + {1 \over 2}  
[(M^{\dagger} M)^{-1/2}]^{\bar j}_{\bar i}  (F^{\bar M})^{\bar i}_i
 (F^M)^{i}_{\bar j} + {g_X^2 \over 2} D_X^2 \ , \label{m20}   
\eea
where $\mu^2 = {3 C_{N_c} / k^2 (T+{\bar T})}$ is a 
mass scale determined by the T-modulus
vev. The new feature of (\ref{m20}) is that $k$ and consequently
the FI term can have both signs, whereas in the effective
heterotic string framework worked out in \cite{bd}, the FI term
had only one possible sign. 

In the limit $\Lambda << \mu$, the vacuum structure and the pattern of
supersymmetry breaking in the two cases of $k$ positive and
negative are vastly different.

i) $k > 0$. In this case the vacuum can be determined as in
\cite{bd}, where it was analysed for arbitrary $N_f < N_c$ and
arbitrary $q + {\bar q}~>~0~$ charges. Keeping one mass parameter 
$m_i^{\bar j} = m \delta_i^{\bar j}$,
we find, to the lowest orders in the parameter $\epsilon$ defined by
\be
\epsilon \equiv {M_0 \over k \mu^2} = 
({\Lambda \over \sqrt{k} \mu})^{3 N_c - N_f \over N_c}
\left[ {m \over M_P} ({\sqrt{k} \mu \over M_P})^{q+{\bar q}-1}
\right]^{N_f-N_c \over N_c} \ , \label{m21} 
\ee
a hierarchically small scale of supersymmetry breaking 
\bea 
&& \langle |\phi|^2 \rangle
= k \mu^2 \left[ 1+ \epsilon N_f (q+{\bar q}) \right] \ , \ 
\langle M \rangle = M_0 \left[ 1- \epsilon (q+{\bar q})^2 
{N_f (N_c-N_f) (2N_c-N_f) \over 2 N_c^2} \right] \ , \nonumber \\
&& g_X^2 \langle D_X \rangle = - \epsilon^2 {\hat m}^2 N_f^2  (q+{\bar q})^2 
\left[ 1 - {N_f \over N_c} (q+{\bar q}) \right] \ , \nonumber \\
&& \langle F_{\phi} \rangle = \epsilon {\hat m} 
\sqrt{k} \mu N_f (q+{\bar q}) \ , \  
\langle F^{\bar M} \rangle = K^{M {\bar M}}  \partial_M W =
- \epsilon^2 {\hat m}  k \mu^2  {N_f (N_c-N_f) \over N_c} (q+{\bar q})^2 
\ , \label{m22} 
\eea
where ${\hat m} \equiv m (\sqrt{k} \mu / M_P)^{q+{\bar q}}$. 

ii) $k < 0$. Here we specifically consider the case $N_c=2$, $N_f=1$
and $q+{\bar q}=1$. In this case we find, to the lowest order in the
parameter $ \epsilon' = [(g^2 + 2 \lambda^2)^4 / 8 \lambda^2 g^{10}]
(\Lambda^2 / |k| \mu^2)^{5}$ , a large
scale of supersymmetry breaking  
 \bea 
&& \langle \phi
\rangle = {(g^2+2 \lambda^2)^2 \over 2 \lambda g^4} {\Lambda^5 \over k^2
  \mu^4 } \left[ 1 + 3 (- g^2 + 14 \lambda^2) \epsilon' \right] \ , \ 
\langle M \rangle = - {g^2 \over g^2 + 2 \lambda^2} k \mu^2 
\left[ 1 - 2 (g^2 + 14 \lambda^2) \epsilon' \right] \ , \ \nonumber \\
&& \langle D_X \rangle \simeq {2 \lambda^2 \over g^2+ 2 \lambda^2} k \mu^2 \
, \ \langle F_{\phi} \rangle \simeq - {\lambda g^2 \over g^2+ 2 \lambda^2}
k \mu^2 \ , \  
\langle F^{\bar M} \rangle \simeq  {g^2 + 2 \lambda^2 \over g^2}
{\Lambda^5 \over k \mu^2 M_P^2} \ . \label{m23} \eea
Interestingly enough, this second case generate a large scale for 
supersymmetry breaking with large $F_{\phi}$ and 
D-term contributions. At first sight,
a breaking of supersymmetry at a scale larger than the dynamical scale
$\Lambda$ destroys the supersymmetric confining dynamics underlying
the nonperturbative superpotential in (\ref{m04}). However, the breaking
of supersymmetry in the hidden sector is described by the mass splitting in the
``mesonic'' sector, measured by the auxiliary field $F_M$. Its value in
case ii) is  very small and actually the same as in case i), suggesting
that the confining dynamics is still essentially supersymmetric.   
In the case $q+ {\bar q} > 1 $ we expect the D-term contribution
to have further suppressions since the mesons masses come now from
a higher dimensional operator. Within this context, we expect our
general analysis of D-term contributions to supersymmetry breaking
to be of relevance for further studies of phenomenological models
incorporating moduli stabilisation \cite{fnop}.   
In the following section we describe 
string theory realisations based on intersecting brane models leading
precisely to the case $q+ {\bar q} = 1$.  
\section{Intersecting brane string realisation of large D-term supersymmetry breaking}

Even if reasonable from a supergravity point of view, it is not
obvious that a large D-term supersymmetry breaking in string
theory is possible. Indeed, it is well known from the heterotic
string constructions that the presence of Fayet-Iliopoulos terms
triggers vev's for charged fields which break the gauge symmetry
rather than supersymmetry \cite{dsw}. This can presumably be
understood by noticing that the FI terms in the heterotic string
arise at one-loop and therefore, if they would break
supersymmetry, they would be a radiative breaking of supersymmetry
which is known to be very hard to obtain \cite{witten}. It was
suggested in \cite{bd} that at the nonperturbative level, gaugino
condensation in the hidden sector in the presence of an anomalous
$U(1)$ symmetry can break supersymmetry. However as    
in the previous section for case i) introduced there, the induced
D-terms are of the order (or slightly larger) than the $F^2/M_P$
type terms and cannot provide the large contributions we are
advocating in this paper. 

In the Type I or Type II strings, on the other hand, the FI terms
appear generically at tree-level and we can expect the tree-level
supersymmetry breaking to be possible with large D-terms. As we will see,
this will realise case ii) discussed in the previous section. We
present here an explicit example suggesting this is indeed
possible, in the context of intersecting branes Type I orientifold
models or, T-dual equivalently, with internal magnetic fields
\cite{earlier,intersecting}. We discuss also various ingredients
such that supersymmetry breaking to be really possible. We
consider an explicit example, even if it is clear that a large
class of similar models can be constructed. The model is based on
the $Z_2 \times Z_2$ Type I orbifold without discrete torsion with
internal magnetic fields $ H^{(a)}_i = (m^{(a)}_i / v_i
n^{(a)}_i)$ in the torus $T^i$, where $v_i$ are the volumes of the
three torii. The model contains four stacks of D9 branes, each
stack containing four coincident branes. Three of the stacks are
magnetised and the fourth one is non-magnetised, with wrapping
numbers $(m_i^{(a)},n_i^{(a)})$ equal to \bea 
&&M_1 \quad : \quad (m^{(3)}_i,n^{(3)}_i) \quad  \quad = \quad (0,1) 
\quad , \quad  (2,1) \quad , \quad (2,1) \ , \nonumber \\
&& M_2 \quad : \quad
(m^{(1)}_i,n^{(1)}_i) \quad \quad = \quad (2,1) \quad , \quad
(0,1) \quad , \quad (2,1) \ , \nonumber \\
&&M_3 \quad : \quad (m^{(2)}_i,n^{(2)}_i) \quad  \quad = \quad (-2,1) 
\quad , \quad
(2,1) \quad , \quad (0,1) \ , \nonumber \\
&&M_4 \quad : \quad(m^{(4)}_i,n^{(4)}_i) \quad  \quad = \quad (0,1) 
\quad ,\quad  (0,1) \quad , \quad (0,1) \ . \
\label{i1} \eea 
The fluxes on $M_1$ and $M_2$ generate lower dimensional anti-brane
like charges whereas  the fluxes on $M_3$ generate lower dimensional
brane like charges. 
The RR tadpole conditions for the $Z_2 \times Z_2$
orbifold without discrete torsion with $(\epsilon_1, \epsilon_2
,\epsilon_3) = (-1,-1, 1) $ are given by \bea && \sum_a M_a
n^{(a)}_1  n^{(a)}_2 n^{(a)}_3 = 16 \quad , \quad \sum_a M_a
n^{(a)}_1  m^{(a)}_2 m^{(a)}_3 = 16  \ , \nonumber \\
&& \sum_a M_a  m^{(a)}_1  n^{(a)}_2 m^{(a)}_3 = 16 \quad , \quad
\sum_a M_a  m^{(a)}_1  m^{(a)}_2 n^{(a)}_3 = - 16 \ \ . \label{i2}
\eea The massless spectrum in this class of models is determined
by the intersection numbers \bea && I_{ab} = \prod_{ab} (n^{(a)}_i
m^{(b)}_i - m^{(a)}_i n^{(b)}_i ) \quad , \quad I_{ab} =
\prod_{ab'} (n^{(a)}_i m^{(b)}_i + m^{(a)}_i n^{(b)}_i ) \ ,
\nonumber \\
&& I_{aO} = 8 (m^{(a)}_i m^{(a)}_i m^{(a)}_i + m^{(a)}_i n^{(a)}_i
n^{(a)}_i + n^{(a)}_i m^{(a)}_i n^{(a)}_i - n^{(a)}_i n^{(a)}_i
m^{(a)}_i) \ . \label{i02}
 \eea The
contribution of the four stacks of branes to the RR tadpole
conditions with wrapping numbers (\ref{i1}) precisely satisfy
(\ref{i2}) when $M_1 = M_2 = M_3 = M_4 = 4$. The gauge group of
this model is $U(2)^3 \otimes SO(4)$. The model was chosen such
that the chiral massless spectrum, determined by the intersection
numbers, to contain only strings stretched between different
stacks of branes. More precisely, by defining $M_i = 2 p_i$, it is
given by  
\bea
&& \phi_{1 ,{\bar a} b}^i \quad : \quad 16 \ \times \ ( {\bf {\bar p}_2, p_3}
) \ , \ \phi_{2 , a c}^j \quad : \quad 16 \ \times \ ( {\bf p_1, p_2} ) 
\ , \nonumber \\
&& \phi_{3 ,{\bar b} {\bar c}}^k \quad : \quad 16 \times \ ( {\bf
{\bar p}_1, {\bar p}_3} ) \ , \label{i3} \eea where the
multiplicity of $16$ in each sector comes from the intersection
numbers of various branes, whereas all other charged states are
non-chiral and will get a mass. There are mixed $U(1)_a \otimes
U(1)_b^2$ and $U(1)_a \otimes SU(p_b)^2$ gauge anomalies in the
model, easily computable from the massless spectrum. 
\be
C_{ab} \ = \ {1 \over 4 \pi^2} \ Tr (X_a X_b^2) \ = \ {2^4 \over 4 \pi^2}
\ Tr (X_a T_b^2) \ = \ {2^8 \over 4 \pi^2}  \pmatrix{\displaystyle
   0  &
   \displaystyle  1 &   \displaystyle -1 \cr
   \displaystyle 1  &
   \displaystyle 0 &  \displaystyle -1 \cr
 \displaystyle -1  &
   \displaystyle 1 &  \displaystyle 0 \cr} \ , 
\label{i03}
\ee
where $X_a$, $a=1,2,3$ are the $U(1)_a$ gauge factors, whereas $T_a$
are the nonabelian generators.  
They are
taken care by axionic couplings of the type $\Theta_a F^a \wedge
F^a$, where $\Theta_a = Im \ f_a$,
where the gauge kinetic functions are given by \be f_a = \prod_i n^{(a)}_i  \ S -
n^{(a)}_1 m^{(a)}_2 m^{(a)}_3  \ T_1 - m^{(a)}_1 n^{(a)}_2
m^{(a)}_3  \ T_2 - m^{(a)}_1 m^{(a)}_2 n^{(a)}_3  \ T_3
\label{i4} \ee and where $Im \ S, Im \ T_i$ are the axion-dilaton
and the three axions associated to the three internal tori. In our
concrete example above, the gauge kinetic functions are explicitly
$f_1 = S-4 T_1$ , $ f_2 = S- 4 T_2 $, $f_3 = S + 4 T_3$. 
The mixed gauge anomalies are taken care
by the nonlinear gauge transformations \bea && \delta \ Im \ T_1 =
{16 \over \pi^2} \ (-\alpha_2 + \alpha_3) \quad , \quad \delta \ 
Im \ T_2 = - {16 \over \pi^2} \ 
(\alpha_1 + \alpha_3) \nonumber
\\ && \delta \ Im \ T_3 = - {16 \over \pi^2} \ 
(\alpha_1 + \alpha_2) \label{i5} \ , \eea
where $\alpha_a$ are the gauge transformation parameters for the
$U(1)_a$ factors. The K{\"a}hler potential contains the terms
\be - \ln \ [T_1 + {\bar T}_1 + {16 \over \pi^2} \ (V_2-V_3)] - \ln \ 
[T_2 + {\bar T}_2 + {16 \over \pi^2} \ (V_1+V_3)] - 
\ln \ (T_3 + {\bar T}_3 + {16 \over \pi^2} \ (V_1+ V_2)] \ee 
which generate FI terms in the effective field theory
\be
\xi_1 = - {16 \over \pi^2} ({1 \over t_2} + {1 \over t_3}) \ , \ 
\xi_2 = - {16 \over \pi^2} ({1 \over t_1} + {1 \over t_3}) \ , \
\xi_3 =  {16 \over \pi^2} ({1 \over t_1} - {1 \over t_2}) \ . \label{i05}
\ee
 The Fayet-Iliopoulos terms can be more generally be written in terms of magnetic fluxes
 as \be \xi_a \ \sim \ H^{(a)}_1 +
H^{(a)}_2 + H^{(a)}_3 - H^{(a)}_1 H^{(a)}_2 H^{(a)}_3 \ \label{i5a}
\ee and in this model they satisfy the sum rule $\xi_1 - \xi_2 -
\xi_3 = 0$. The D-terms on the U(1) factors of each $U(2) = U(1)
\otimes SU(2)$ stack are given by \bea && D_1 \ = \ 
\sum_{i,{\bar a},b} |\phi_{2 ,{\bar a} b}^i|^2 -
\sum_{j, a,c} |\phi_{3 , a c}^j|^2 + \xi_1 \ , \nonumber \\
&& D_2 \ = \  -\sum_{i,{\bar a},b} |\phi_{1 ,{\bar a} b}^i|^2 +
\sum_{k, {\bar b},{\bar c}} |\phi_{2 ,{\bar b} {\bar c}}^k|^2 + \xi_2 \ , \nonumber \\
&& D_3 \ = \  \sum_{j, a,c} |\phi_{1 , a c}^j|^2 - \sum_{k, {\bar
b},{\bar c}} |\phi_{3 ,{\bar b} {\bar c}}^k|^2 + \xi_3 \ ,
\label{i6} \eea and satisfy also the same rule \be D_1 - D_2 - D_3
= \xi_1-\xi_2-\xi_3 = 0 \ . \label{i7} \ee We believe that the interpretation of the
sum rule (\ref{i7}) is that the D-branes tend to recombine by
condensing the by-fundamental fields (\ref{i3}) and to provide a
supersymmetric vacuum $D_a=0$.
 Notice however that the model has  renormalizable
superpotential terms \be W \ = \ \lambda_{ijk} \ Tr \ (\phi_1^i
\phi_2^j \phi_3^k ) \ , \label{i4a} \ee (where the trace in the
gauge group space), which has a geometrical interpretation of
Yukawa couplings connecting 3-fields forming a triangle in each
compact torus, analogously to models of Yukawa couplings studied
in \cite{yukawas}. The number of these Yukawas are related, as
usual, to the number of D-flat directions. The Yukawa couplings
are also field dependent and depend on complex structure moduli.
In the presence of the Yukawa couplings (\ref{i4a}), the tachyonic
instabilities typically related to the brane recombination process
can be removed, there is
generically a geometrical obstruction and the D-brane
recombination is not generically the most favorable process. 
The FI terms are K{\"a}hler moduli dependent and they can
(perturbatively ) vanish for particular points in the K{\"a}hler
moduli space, which will always be dynamically preferred. In order
to avoid this phenomenon, nonperturbative effects have to be
invoked, for example gaugino condensation on
the D9 branes, according to the discussion in the previous section. 
The perturbative nature of the superpotential terms (\ref{i4a}), to be
interpreted as meson masses in case ii) of the previous section,
generate large D-term contributions . Cosmological
constant cancellation for large D-terms in the TeV range gravitino
mass ask for intermediate values of FI terms and/or very small
$U(1)$ gauge couplings. So for large scale of supersymmetry breaking, 
cosmological constant is generically hard to cancel unless FI terms
are much smaller than the Planck scale. Whereas this was not the case
in the nonperturbative model of the previous section, the
more general formula (\ref{i5a}) suggests that it is possible that
$\xi_a << H_i^{(a)}$ by tuning the magnetic fluxes, in the spirit
of landscape models \cite{bp,split,ad,kklt,ddg}. 
 
In order to make connection with the field theory model of the
previous section, notice that if dynamics picks up an overall Kahler
modulus $T_1=T_2=T_3 = T$, then $\xi_3=0$ and there are two remaining
anomalous abelian factors $U(1)_1$ , $U(1)_2$. In the following we
discuss in some more detail a simplified model along these lines.     

\subsection{From intersecting brane models to nonperturbative moduli 
stabilisation}
In order to stabilise all moduli, in section 4 we used nonpeturbative 
effects on an asymptotically-free gauge group.  The
explicit string example discussed previously has no asymptotically
gauge factor, but we do not expect this result to be generic.   In
the following, we consider a model, similar to the explicit string
example presented in the previous section but containing an
asymptotically-free gauge factor. It is also further simplified in
order to allow a simple analysis and suited to generate large
D-terms at the minimum, comparable and in some regions of the
parameter space dominant with respect to the F-terms.

The field content and gauge structure are summarized as follows.
The model has a gauge group $SU(N)\otimes U(1)^2$, with the chiral
superfield content \be  \phi_1^i : (N,1,0) \quad , \quad
\phi_{2,{\bar j}} : ({\bar N},0,-1) \quad ,  
\quad \phi_3 : (1,-1,+1) \ , \label{m1}
 \ee
where $i, {\bar j} = 1 \cdots N_f$ and the notation for charges
and representations are transparent in (\ref{m1}). The model is
therefore similar to the explicit intersecting brane model of the
previous section, but slightly adapted for our purposes. There is
a magnetic flux pattern which does lead to the spectrum above,
which by itself does not saturate the RR tadpole conditions. This
can be cured by adding additional branes or by considering other
orbifolds and/or additional antisymmetric field backgrounds. The
$SU(N)$ plays the role of a hidden sector SYM gauge group with
$N_f$ flavors, which condenses in the IR. The composite objects
\be M^i_{\bar j} = \phi_1^{ia} \phi_{2,{\bar j}}^{\bar a}
\label{m2} , \ee  where $a$ is an index in the fundamental of
$SU(N)$, are the mesons used in constructing the effective action
of the theory. For simplicity we consider in the following only
the overall K{\"a}hler modulus $T$, whereas keeping all of them would
ask for stabilisation a more complicated dynamics, for example
several gaugino condensates. Consistently with the cancellation of
the mixed gauge anomalies, the gauge kinetic function on the
condensing gauge group is $f_{SU(N)} = S \pm N_f T$, where the $+$ ($-$)
signs correspond to a hidden sector with positive (negative) product
of magnetic fluxes in two torii. Similar to the
explicit intersecting brane model, $T$ transforms under gauge
transformations as \be \delta T = \pm {1 \over 4 \pi^2} \ 
(\Lambda_2 - \Lambda_3) \ .\label{m020} 
\ee 
This can also be directly checked by computing the
mixed gauge anomalies \be  \ U(1)_2 \otimes SU(N)^2 \ : \ {N_f \over 4 \pi^2} 
\quad , \ U(1)_3 \otimes SU(N)^2 \ : \
- {N_f \over 4 \pi^2} \ , \ee 
which are precisely canceled by the nonlinear gauge
transformation of the axion $Im T$ (\ref{m020}).   
 In order to write the K{\"a}hler potential, first of all we place ourselves on the $SU(N)$ flat direction
 $\langle \phi_1^{ia} \rangle = \langle \phi_{2,{\bar j}}^{\bar a} \rangle$.
 Similarly to the KKLT proposal, we could first integrate out the dilaton 
and the complex structure moduli. In doing this, for $N_f < N$,
 we find the effective superpotential and K{\"a}hler potential 
\bea && W = W_0+ (N-N_f) A \left[ {e^{\mp  8 \pi^2 N_f T } \over det M_{\bar
j}^i}
 \right]^{1 \over N-N_f} + \lambda_{i {\bar j}} \phi_3 M^i_{\bar j} \ , 
\nonumber \\
&& K = - 3 \ln \ \bigl[ T + {\bar T} \pm {V_3 - V_2 \over 4 \pi^2} - 2 Tr ({\bar M}
 M)^{1 / 2} - {\bar \phi}_3  \phi_3 \bigr]
\ , \label{m4} \eea where the constant $W_0$ depend on the details
of the three-form fluxes, the K{\"a}hler potential was computed
in the weakly coupled regime of the $SU(N)$ flat direction and where
$A = \exp \{- 8 \pi^2 S_0/(N-N_f) \}$. 

The reader will notice that this model reassembles closely the
model worked out in Section 4. The mass term for the mesons in
(\ref{m4}) is actually provided, in the intersecting brane
realisation of the previous section, by the Yukawa coupling
(\ref{i4a}).
 In analogy with the explicit intersecting brane model, there is a
 constraint equation
 \be
D_2 - D_3 \ = \ \xi_2 - \xi_3 \ = \ 0 \ , \label{m5}
 \ee
 where
 \be
D_2 \ = \ Tr ({\bar M} M)^{1 / 2} - |\phi_3|^2 \pm  {3 \over 4 \pi^2 (T + {\bar
T})} \ ,
 \ee
which signals the presence of a flat direction, allowing the
presence of the perturbative superpotential providing the meson
mass term. Indeed, since there are now two D-flatness conditions and
two charged fields, the existence of the flat direction (\ref{m5})
is needed in order to write the meson mass term, the last term in 
the superpotential (\ref{m4}). The two signs in the expressions above
correspond to the two cases $k >0$ and $k <0$ in Section 4, the
second case realising the high scale supersymmetry breaking with large D-terms. 

\section{Summary and Outlook}
In the present work, we have initiated  a program to study
in a general manner the implications of the large D-terms in a 
supergravity on soft supersymmetry breaking parameters. These 
terms can come from an anomalous $U(1)$ flavour model as has been
noted in the past or tree level FI terms in an intersecting
D-brane model. We have shown that explicit models based on 
intersecting D-branes can be constructed giving rise to large D-terms. 
Irrespective of the source, we have studied the
implications of these terms on the soft parameters. The mass squared
terms are the most affected with contributions linear in D. 
However the charges of the matter fields under
the anomalous $U(1)$ can crucially determine the actual impact. 
The $B_\mu$ and $A$ terms also receive corrections though they
are not significantly modified in terms of magnitude. As an application
we have shown that split supersymmetry can be realised with  
specific choices of the superpotential and K{\"a}hler potential. 

Particular examples with the large $D$ contributions are
string models of supersymmetry breaking, in particular in
Type I string orientifolds. We have not addressed in detail the
phenomenological signatures and constraints on the parameter space
within this class of the models. Such a study could be taken in 
conjunction with a proper flavour model {\`a} la Froggatt-Nielsen.
This could be then confronted with low energy data from accelerators,
dark matter physics constraints and flavour physics.

The issue of moduli stabilisation has been receiving increasing
attention in the recent years. Applications to the soft masses
have also been recently addressed\cite{fnop,quevedo}. Here we have
revisited the issue of moduli stabilisation using non-perturbative
gaugino condensation in type I orientifolds with internal magnetic
fluxes which generate large D-terms. By a suitable
choice of the fluxes which fix the sign of the FI term, we 
have shown that it is possible
to stabilize moduli and generate large D-terms. 


\section*{Acknowledgments}
\noindent We wish to thank  S. Lavignac, Y. Mambrini, S. Pokorski
and Carlos Savoy for useful discussions.

\noindent
This work is supported in part by the CNRS PICS no. 2530 and 3059, INTAS grant
03-51-6346, the RTN grants MRTN-CT-2004-503369, MRTN-CT-2004-005104 and
by a European Union Excellence Grant, MEXT-CT-2003-509661.
E.D. thanks the Univ. of Tucson, Arizona for warm hospitality while
completing this work. 
SKV is also supported by Indo-French Centre for Promotion of
Advanced Research (CEFIPRA) project No:  2904-2 `Brane World Phenomenology'.


\begin{thebibliography}{99}

\bibitem{carlos}
R.~Barbieri, S.~Ferrara and C.~A.~Savoy,
Phys.\ Lett.\ B {\bf 119}, 343 (1982);
%
A.~H.~Chamseddine, R.~Arnowitt and P.~Nath,
Phys.\ Rev.\ Lett.\  {\bf 49}, 970 (1982);
%
L.~J.~Hall, J.~Lykken and S.~Weinberg,
Phys.\ Rev.\ D {\bf 27} (1983) 2359.

\bibitem{nilles}
H.~P.~Nilles,
Phys.\ Rept.\  {\bf 110} (1984) 1.

\bibitem{weldonsoni}
S.~K.~Soni and H.~A.~Weldon,
Phys.\ Lett.\ B {\bf 126}, 215 (1983).
%
\bibitem{kaplunovskylouis}
V.~S.~Kaplunovsky and J.~Louis,
Phys.\ Lett.\ B {\bf 306}, 269 (1993)
[arXiv:hep-th/9303040].

\bibitem{ibanez}
A.~Brignole, L.~E.~Ibanez and C.~Munoz,
  Nucl.\ Phys.\ B {\bf 422}, 125 (1994)
  [Erratum-ibid.\ B {\bf 436}, 747 (1995)]
  [arXiv:hep-ph/9308271]~;
A.~Brignole, L.~E.~Ibanez and C.~Munoz,
arXiv:hep-ph/9707209.
%
%
\bibitem{ibaneztypeI}
L.~E.~Ibanez, C.~Munoz and S.~Rigolin,
Nucl.\ Phys.\ B {\bf 553}, 43 (1999)
[arXiv:hep-ph/9812397].
%
\bibitem{dps}
E.~Dudas, S.~Pokorski and C.~A.~Savoy,
Phys.\ Lett.\ B {\bf 369}, 255 (1996)
[arXiv:hep-ph/9509410;
E.~Dudas, C.~Grojean, S.~Pokorski and C.~A.~Savoy,
Nucl.\ Phys.\ B {\bf 481}, 85 (1996)
[arXiv:hep-ph/9606383];
Y.~Kawamura and T.~Kobayashi,
Phys.\ Lett.\ B {\bf 375}, 141 (1996)
[Erratum-ibid.\ B {\bf 388}, 867 (1996)]
[arXiv:hep-ph/9601365] and
Phys.\ Rev.\ D {\bf 56}, 3844 (1997)
[arXiv:hep-ph/9608233].

\bibitem{bd}
P.~Binetruy and E.~Dudas,
Phys.\ Lett.\ B {\bf 389}, 503 (1996)
[arXiv:hep-th/9607172];
G.~R.~Dvali and A.~Pomarol,
Phys.\ Rev.\ Lett.\  {\bf 77}, 3728 (1996)
[arXiv:hep-ph/9607383].

%
\bibitem{bp}
S.~Weinberg,
Phys.\ Rev.\ Lett.\  {\bf 59}, 2607 (1987);
%
R.~Bousso and J.~Polchinski,
JHEP {\bf 0006}, 006 (2000)
[arXiv:hep-th/0004134].
%
\bibitem{split}
N.~Arkani-Hamed and S.~Dimopoulos,
arXiv:hep-th/0405159;
G.~F.~Giudice and A.~Romanino,
Nucl.\ Phys.\ B {\bf 699}, 65 (2004)
[arXiv:hep-ph/0406088]~.

\bibitem{splitpheno}
An incomplete list of references:
L.~Anchordoqui, H.~Goldberg and C.~Nunez,
Phys.\ Rev.\ D {\bf 71}, 065014 (2005)
[arXiv:hep-ph/0408284];
%
N.~Arkani-Hamed, S.~Dimopoulos, G.~F.~Giudice and A.~Romanino,
arXiv:hep-ph/0409232;
%
D.~A.~Demir,
arXiv:hep-ph/0410056;
%
U.~Sarkar,
arXiv:hep-ph/0410104;
%
R.~Allahverdi, A.~Jokinen and A.~Mazumdar,
Phys.\ Rev.\ D {\bf 71}, 043505 (2005)
[arXiv:hep-ph/0410169];
%
E.~J.~Chun and S.~C.~Park,
JHEP {\bf 0501}, 009 (2005)
[arXiv:hep-ph/0410242];
%
B.~Bajc and G.~Senjanovic,
Phys.\ Lett.\ B {\bf 610}, 80 (2005)
[arXiv:hep-ph/0411193];
%
A.~Arvanitaki and P.~W.~Graham,
 arXiv:hep-ph/0411376;
%
A.~Masiero, S.~Profumo and P.~Ullio,
Nucl.\ Phys.\ B {\bf 712}, 86 (2005)
[arXiv:hep-ph/0412058];
%
M.~A.~Diaz and P.~F.~Perez,
J.\ Phys.\ G {\bf 31}, 563 (2005)
[arXiv:hep-ph/0412066];
%
L.~Senatore,
Phys.\ Rev.\ D {\bf 71}, 103510 (2005)
[arXiv:hep-ph/0412103];
%
A.~Datta and X.~Zhang,
arXiv:hep-ph/0412255;
%
M.~Beccaria, F.~M.~Renard and C.~Verzegnassi,
Phys.\ Rev.\ D {\bf 71}, 093008 (2005)
[arXiv:hep-ph/0412257];
%
P.~C.~Schuster,
arXiv:hep-ph/0412263;
%
S.~P.~Martin, K.~Tobe and J.~D.~Wells,
Phys.\ Rev.\ D {\bf 71}, 073014 (2005)
[arXiv:hep-ph/0412424];
%
C.~H.~Chen and C.~Q.~Geng,
arXiv:hep-ph/0501001;
%
M.~Drees,
arXiv:hep-ph/0501106;
%
S.~Kasuya and F.~Takahashi,
arXiv:hep-ph/0501240;
%
K.~Huitu, J.~Laamanen, P.~Roy and S.~Roy,
arXiv:hep-ph/0502052;
%
G.~Marandella, C.~Schappacher and A.~Strumia,
Nucl.\ Phys.\ B {\bf 715}, 173 (2005)
[arXiv:hep-ph/0502095];
%
N.~Haba and N.~Okada,
arXiv:hep-ph/0502213;
%
C.~H.~Chen and C.~Q.~Geng,
arXiv:hep-ph/0502246;
%
B.~Dutta and Y.~Mimura,
arXiv:hep-ph/0503052;
%
D.~Chang, W.~F.~Chang and W.~Y.~Keung,
Phys.\ Rev.\ D {\bf 71}, 076006 (2005)
[arXiv:hep-ph/0503055];
%
N.~G.~Deshpande and J.~Jiang,
Phys.\ Lett.\ B {\bf 615}, 111 (2005)
[arXiv:hep-ph/0503116];
%
A.~Ibarra,
arXiv:hep-ph/0503160;
%
B.~Mukhopadhyaya and S.~SenGupta,
arXiv:hep-ph/0503167;
%
M.~Toharia and J.~D.~Wells,
arXiv:hep-ph/0503175;
%
R.~N.~Mohapatra, M.~K.~Parida and G.~Rajasekaran,
arXiv:hep-ph/0504236;
%
J.~G.~Gonzalez, S.~Reucroft and J.~Swain,
arXiv:hep-ph/0504260.
%
\bibitem{dv}
 E.~Dudas and S.~K.~Vempati,
  arXiv:hep-ph/0506029.


%

\bibitem{wessbagger}
J.~Wess and J.~Bagger,
``Supersymmetry and supergravity,''
 Princeton, USA: Univ. Pr. (1992).
%
\bibitem{fabio}
S.~Ferrara, C.~Kounnas and F.~Zwirner,
Nucl.\ Phys.\ B {\bf 429}, 589 (1994)
[Erratum-ibid.\ B {\bf 433}, 255 (1995)]
[arXiv:hep-th/9405188].
%

\bibitem{amsb1}
L.~Randall and R.~Sundrum,
Nucl.\ Phys.\ B {\bf 557}, 79 (1999)
[arXiv:hep-th/9810155];
%
G.~F.~Giudice, M.~A.~Luty, H.~Murayama and R.~Rattazzi,
JHEP {\bf 9812}, 027 (1998)
[arXiv:hep-ph/9810442].
%

\bibitem{amsb2}
J.~A.~Bagger, T.~Moroi and E.~Poppitz,
JHEP {\bf 0004}, 009 (2000)
[arXiv:hep-th/9911029];
%
P.~Binetruy, M.~K.~Gaillard and B.~D.~Nelson,
Nucl.\ Phys.\ B {\bf 604}, 32 (2001)
[arXiv:hep-ph/0011081].
%

\bibitem{gm}
G.~F.~Giudice and A.~Masiero,
Phys.\ Lett.\ B {\bf 206}, 480 (1988).
%
\bibitem{freedman}
  D.~Z.~Freedman,
 Phys.\ Rev.\ D {\bf 15}, 1173 (1977);
%
P.~Binetruy, G.~Dvali, R.~Kallosh and A.~Van Proeyen,
 Class.\ Quant.\ Grav.\  {\bf 21}, 3137 (2004)
 [arXiv:hep-th/0402046].

\bibitem{ir}
L.~E.~Ibanez and G.~G.~Ross,
  Phys.\ Lett.\ B {\bf 332}, 100 (1994)
  [arXiv:hep-ph/9403338];
P.~Binetruy and P.~Ramond,
  Phys.\ Lett.\ B {\bf 350}, 49 (1995)
  [arXiv:hep-ph/9412385];
E.~Dudas, S.~Pokorski and C.~A.~Savoy,
  Phys.\ Lett.\ B {\bf 356}, 45 (1995)
  [arXiv:hep-ph/9504292].
Y.~Nir,
  Phys.\ Lett.\ B {\bf 354}, 107 (1995)
  [arXiv:hep-ph/9504312].

\bibitem{ns}
Y.~Nir and N.~Seiberg,
  Phys.\ Lett.\ B {\bf 309}, 337 (1993)
  [arXiv:hep-ph/9304307];
%
M.~Leurer, Y.~Nir and N.~Seiberg,
Nucl.\ Phys.\ B {\bf 420}, 468 (1994)
[arXiv:hep-ph/9310320].
%

\bibitem{wells}
J.~D.~Wells,
Phys.\ Rev.\ D {\bf 71}, 015013 (2005)
[arXiv:hep-ph/0411041].


\bibitem{ad}
 I.~Antoniadis and S.~Dimopoulos,
  arXiv:hep-th/0411032;
see also C.~Kokorelis,
  arXiv:hep-th/0406258.

\bibitem{nelson}
P.~J.~Fox, A.~E.~Nelson and N.~Weiner,
  JHEP {\bf 0208}, 035 (2002)
  [arXiv:hep-ph/0206096];
 Z.~Chacko, P.~J.~Fox and H.~Murayama,
  Nucl.\ Phys.\ B {\bf 706}, 53 (2005)
  [arXiv:hep-ph/0406142].


\bibitem{cremmer}
E.~Cremmer, S.~Ferrara, L.~Girardello and A.~Van Proeyen,
Nucl.\ Phys.\ B {\bf 212} (1983) 413.
%

\bibitem{dvalipomarol}
G.~R.~Dvali and A.~Pomarol,
Phys.\ Lett.\ B {\bf 410}, 160 (1997)
[arXiv:hep-ph/9706429].

\bibitem{kn}
B.~Kors and P.~Nath,
arXiv:hep-th/0411201;
K.~S.~Babu, T.~Enkhbat and B.~Mukhopadhyaya,
arXiv:hep-ph/0501079.

%
\bibitem{quevedoreview}
For reviews, see for e.g,
  F.~Quevedo,
  arXiv:hep-th/9603074 ;
E.~Dudas,
Class.\ Quant.\ Grav.\  {\bf 17}, R41 (2000)
[arXiv:hep-ph/0006190];
E.~Kiritsis,
Fortsch.\ Phys.\  {\bf 52}, 200 (2004) [arXiv:hep-th/0310001].

\bibitem{dsw}
 M.~Dine, N.~Seiberg and E.~Witten,
  Nucl.\ Phys.\ B {\bf 289}, 589 (1987);
J.~J.~Atick, L.~J.~Dixon and A.~Sen,
  Nucl.\ Phys.\ B {\bf 292}, 109 (1987).

\bibitem{witten}
E.~Witten,
  Nucl.\ Phys.\ B {\bf 202}, 253 (1982).

\bibitem{ads}
T.~R.~Taylor, G.~Veneziano and S.~Yankielowicz,
  Nucl.\ Phys.\ B {\bf 218}, 493 (1983);
I.~Affleck, M.~Dine and N.~Seiberg,
  Nucl.\ Phys.\ B {\bf 241}, 493 (1984).

\bibitem{earlier}
C.~Bachas,
arXiv:hep-th/9503030;
M.~Berkooz, M.~R.~Douglas and R.~G.~Leigh,
Nucl.\ Phys.\ B {\bf 480} (1996) 265 [arXiv:hep-th/9606139];
N.~Ohta and P.~K.~Townsend,
Phys.\ Lett.\ B {\bf 418} (1998) 77 [arXiv:hep-th/9710129];

\bibitem{intersecting}
R.~Blumenhagen, L.~Goerlich, B.~Kors and D.~Lust,
JHEP {\bf 0010} (2000) 006 [arXiv:hep-th/0007024];
C.~Angelantonj, I.~Antoniadis, E.~Dudas and A.~Sagnotti,
Phys.\ Lett.\ B {\bf 489} (2000) 223 [arXiv:hep-th/0007090]~;
G.~Aldazabal, S.~Franco, L.~E.~Ibanez, R.~Rabadan and
A.~M.~Uranga,
JHEP {\bf 0102} (2001) 047 [arXiv:hep-ph/0011132];
M.~Cvetic, G.~Shiu and A.~M.~Uranga,
Nucl.\ Phys.\ B {\bf 615} (2001) 3 [arXiv:hep-th/0107166]
and \ Phys.\ Rev.\ Lett.\  {\bf 87}, 201801 (2001)
[arXiv:hep-th/0107143].

\bibitem{yukawas}
 D.~Cremades, L.~E.~Ibanez and F.~Marchesano,
  JHEP {\bf 0307}, 038 (2003)
  [arXiv:hep-th/0302105];
M.~Cvetic and I.~Papadimitriou,
  Phys.\ Rev.\ D {\bf 68}, 046001 (2003)
  [Erratum-ibid.\ D {\bf 70}, 029903 (2004)]
  [arXiv:hep-th/0303083];
 S.~A.~Abel and A.~W.~Owen,
  Nucl.\ Phys.\ B {\bf 663}, 197 (2003)
  [arXiv:hep-th/0303124].

\bibitem{ddg}
K.~R.~Dienes, E.~Dudas and T.~Gherghetta,
arXiv:hep-th/0412185~;
N.~Arkani-Hamed, S.~Dimopoulos and S.~Kachru,
  arXiv:hep-th/0501082.

\bibitem{kklt}
S.~Kachru, R.~Kallosh, A.~Linde and S.~P.~Trivedi,
  Phys.\ Rev.\ D {\bf 68}, 046005 (2003)
  [arXiv:hep-th/0301240].

\bibitem{bkq}
C.~P.~Burgess, R.~Kallosh and F.~Quevedo,
  JHEP {\bf 0310}, 056 (2003)
  [arXiv:hep-th/0309187].

\bibitem{fnop}
K.~Choi, A.~Falkowski, H.~P.~Nilles, M.~Olechowski and S.~Pokorski,
  JHEP {\bf 0411}, 076 (2004)
  [arXiv:hep-th/0411066]; 
K.~Choi, A.~Falkowski, H.~P.~Nilles and M.~Olechowski,
  Nucl.\ Phys.\ B {\bf 718}, 113 (2005)
  [arXiv:hep-th/0503216]; 
 K.~Choi, K.~S.~Jeong and K.~i.~Okumura,
  arXiv:hep-ph/0504037. 

\bibitem{quevedo}
 J.~P.~Conlon, F.~Quevedo and K.~Suruliz,
 arXiv:hep-th/0505076.


\end{thebibliography}
\end{document}